\documentclass{aa}
\usepackage{psfig}
\def\teff{$T\rm_{eff}$}
\def\lambo{$\lambda$ Boo }
\usepackage{graphics}
\begin{document}
\thesaurus{08(08.01.3;08.01.1;08.03.2)}

\title{How many $\lambda$ Boo stars are binaries ?}
\thanks{Partly based on data from the ESA Hipparcos astrometry satellite.}
\author{Rosanna \,Faraggiana\inst{1},
\and Piercarlo\,Bonifacio\inst{2}}
\offprints{R. Faraggiana}
\institute{Dipartimento di Astronomia, Universit\`a degli Studi di Trieste,
Via G.B.Tiepolo 11, I-34131 Trieste, Italy
\and Osservatorio Astronomico di Trieste,Via G.B.Tiepolo 11,
I-34131 Trieste, Italy}
\mail{faraggiana@ts.astro.it}
\date{received .../Accepted...}
\
\maketitle

\begin{abstract}
In the attempt to shed new light on the $\lambda$ Boo
phenomenon we analyzed the astrometric, photometric and
spectroscopic characteristics of stars out of a list of recently selected
$\lambda$ Boo candidates.

We show that the class is still ill-defined and
discuss the possibility that some, if not most stars
presently classified as $\lambda$ Boo, are
in fact binary pairs and that peculiar abundances may 
not correspond to actual values if
the average values of the atmospheric parameters \teff~and log\,g are
assumed and the effect of veiling is not taken into account.

\end{abstract}
\keywords{08.01.3 Stars: atmospheres - 08.01.1 Stars: abundances -
08.03.2 Stars: Chemically Peculiar }

\section{Introduction}

The class of \lambo stars still presents poorly defined characteristics,
and this more than 50 years after the discovery of the prototype member 
\lambo 
by Morgan et al (1943) who noted the abnormally weak metal lines of this
A0 dwarf star.

The properties that should define a \lambo star are not clearly established;
the proposed spectroscopic criteria are usually based on the weakness of metal
lines, especially of the Mg II 4481, compared with what
is expected from the hydrogen line type, while C, N, O and S have nearly solar
abundances. The kinematic behaviour should allow to distinguish these stars
from the metal poor A-type Horizontal Branch stars.
Moderate to high projected rotational velocities are usually found among
\lambo stars, although some exceptions have been recently identified 
(e.g. HD 64491 and HD 74873 selected by Paunzen \& Gray 1997).

The result of the vague definitions of these non-evolved metal 
underabundant stars
is well reflected by the variety of opinions existing at present about the
members of this class.
The metal abundances obtained up to now reveal a high scatter from star
to star. \par

Details on the evolution with time of the  \lambo definition are summarized
in Faraggiana \& Gerbaldi (1998); the not clearly defined properties of
these stars are, at least partially, responsible of the various hypotheses
proposed to explain the \lambo phenomenon as well as of the uncertainty on the
age attributed to these objects, which spans from that of stars not 
yet 
on the Main Sequence to that of old objects descending from 
contact
binary systems.

The present paper reviews the characteristics of the members of this class
according to recent compilations and discusses the effect of duplicity
on a composite spectrum as  source of misclassification for some
of these candidates.

In a modern astrophysical perspective,
age, position in the HR diagram
and chemical abundances are the key
quantities  which describe a class of stars.
The purpose of introducing a 
class of stars is to help identify  a common underlying
phenomenology.
For \lambo stars the aim is to find
the  common factor which can explain the 
observed chemical peculiarities;
to be meaningful this must explain a statistically significant sample
of stars.
Bearing this in mind, it is clear that any classification scheme
which does not rely on abundance criteria is unlikely to be helpful.
It is probable that as high accuracy abundance data 
accumulate,
we will have to revise our concept of \lambo stars
and probably reach a more physical definition.

\section{The $\lambda$ Boo candidates}

Many stars were classified as \lambo in the past.
The catalogue of Renson et al (1990) includes over 100 candidates.
Many of these turned out to be misclassified and the whole sample
results too heterogeneous.
We  selected
 stars  classified as \lambo in recent papers based on modern
data, hoping to extract a more homogeneous sample.
They should be
considered \lambo candidates, since
for many of them further analysis 
to check whether they match any given \lambo definition is still required.
The candidates we selected, with the exception of three of them,
have been listed in at least one of the following
papers: Abt \& Morrell(1995; hereafter AM), Paunzen et al (1997; hereafter CC),
Gray (1999; hereafter G). 
The exceptions are: HD 290492
and HD 90821 which were classified as \lambo  by Paunzen \&
Gray (1997); HD 105759 for which the G classification is 
unpublished.

Both methods and scope differ among the three papers (i.e. AM,CC,G).  
However,  the degree of reliability of
each of them is difficult to define. 
Promising candidates are found in all lists, although
Gray is probably the most reliable source, because the author 
(Gray \& Garrison 1987, 1989a,1989b; Garrison \& Gray 1994; GG in Table 1)
has classified a large sample of ``normal''
and ``standard'' stars using the same methods.

AM is a study of stellar v $ \sin i$ of 1700 A-type stars of the
Bright Star Catalogue (Hoffleit \& Warren 1994) (BSC). 
On the basis of their available spectra
(photographic  spectra of dispersion
39 \AA $\rm mm^{-1}$)
they give a classification
for each star.  
Some  are classified  as $\lambda$ Boo. 
\footnote{AM classification 
has been criticized  by the referee
(Dr. E. Paunzen) because 
the standards of Morgan et al. (1978) defined for
a dispersion of 125 A mm$^{-1}$ are used, in spite of the higher
dispersion available.
In fact since most of the \lambo stars have $v \sin i > 150$ km/s
a higher dispersion does not improve the results.
However, AM note that, even if this procedure
may in fact lead to an inaccuracy
in the spectral type of 0.27 subclasses
they could see faint lines better.
}

CC, on the contrary, is a paper specifically aimed at \lambo stars.
Their working definition is
``Pop I hydrogen burning A-type stars, which are, except of C, N, O and S,
metal poor''.
The catalogue contains 45 objects and includes stars with a \teff~ as low
as 6500 K. 
The stars should
have the same characteristics of those of $\lambda$ Boo itself and
establish a homogeneous group of $\lambda$ Boo stars.
According to the criteria adopted by the authors, the log g must be consistent
with a Main Sequence
evolutionary status, 
but the photometrically derived log g is less than 3.6 
for 9 of the 45 selected objects. Accurate abundances for some elements 
are available for less than half of these stars and the 
abundances of the key elements C,N,O and S for an even smaller number of
objects. 
Keeping in mind our remarks on the importance
of abundances in  establishing the \lambo status,
this status still has to be confirmed for over one half 
of the stars in CC.

G is also specifically devoted to \lambo
stars.
It is based on his own accurate classification for all but one star (for which
the abundance analysis has been performed by St\"urenburg (1993; hereafter 
St93)). This classification 
is based on the comparison with a set of a previously selected sample of
standard stars with various v$\sin i$ values.
This list must be considered as the most reliable source of
\lambo candidates because it is based on
 a large set of homogeneous data.

Since the \lambo nature rests ultimately on
a peculiar abundance pattern,
only an accurate abundance analysis based on high quality
data and covering a broad spectral range can 
allow to either retain or reject these candidates.

The list assembled in such a way
comprises 89 objects,  9 of which are in common with CC, G and AM;
further 7 further objects are in common between CC and G, but were not observed by AM.
Thus out of 89 candidates there is concordance among different classificators
at most in 16 cases, i.e.  less than 20 \%.
We interpret this poor agreement as evidence of the subjective
classification criteria adopted by each author. \par

We intend to examine the properties of these stars 
 with the aim at investigating:

i) if a homogeneous group can be selected;

ii) if one of the hitherto proposed theories about the origin of the
$\lambda$ Boo phenomenon is able to explain {\it all} the observed
peculiarities of these stars.

\setlength{\tabcolsep}{2pt}
\begin{table*}
\caption{The $\lambda$ Boo candidates.}
\label{tab1}
\begin{tabular}{lccclccccccr}
\hline
~~~HD & CC & G & G+GG classif. & AM & AM classif. &AM  &Sep &$\Delta m$ &Sep &$\Delta m$ &Rem.  \\
& &           &      &    &       &   v$ \sin i$ & $''$ & mag & $''$ & mag                  \\
& &           &      &    &       &               & HIP & HIP & WDS & WDS                  \\
\hline
 \\
\hspace{5.5mm} 3 &   &   & & x & A0Vn($\lambda$Boo)          & 210: & & & & & D \\
\hspace{3mm} 319 & x & x & A1mA2 Vb $\lambda$Boo PHL & & A2Vp(4481wk) & 45 & & &  2.1& 5.1  \\
\hspace{2mm} 2904 &   &   & & x &  A0Vnn($\lambda$Boo)       & 225: \\
\hspace{2mm} 4158$^1$ & x &   & &   & ~~~~~~~--                     &   \\
\hspace{2mm} 5789 &   &   & & x & B9.5Vnn ($\lambda$Boo) & 230:    & & & 7.8 & 0.8 &  \\
\hspace{2mm} 6870 & x &   & &  & ~~~~~~~--                     & &  \\
~~11413 & x & x &  A1 Va $\lambda$Boo PHL & x & A1Vp($\lambda$Boo)  & ...& & & & &  U           \\
~~11503 &   &   & B9.5 IV$^{+}$n & x & A0Vp($\lambda$Boo)n         & 185    & &  &  8.6 & 0.0       \\
~~22470 &   &   & & x & B9.5p ($\lambda$Boo)        &  65    & 0.152  & 1.36        \\
~~23258 &   &   & & x & A0Vp($\lambda$Boo)          & 110           \\
~~23392 &   & x & A0 Va- ($\lambda$Boo) NHL &  & ~~~~~~~--                   &               \\
~~30422 & x & x & A3 Vb $\lambda$Boo PHL &  & A3Vp(4481wk)                & 100           \\
~~30739 &   &   & A0.5 IVn &  x & A0Vp($\lambda$Boo)          & 195           \\
~~31295 & x & x & A0 Va $\lambda$Boo NHL & x & A0Vp($\lambda$Boo)          & 105           \\
~~34787 &   &   & & x & B9.5Vp($\lambda$Boo)        & 200:          \\
~~36726 &   & x & kA0hA5mA0 $\lambda$Boo NHL&  & ~~~~~~~--                   &               \\
~290492$^2$ &   &   &  &                     &           &   & &    &  0.6 &  1.4 &        \\
~294253 & x & x & B9.5 Va ($\lambda$Boo) NHL &  & ~~~~~~~--                    &              \\
~290799 & x & x & A2 Vb $\lambda$Boo PHL &  & ~~~~~~~--                    &              \\
~~38545 & x & x & A2 Va$^{-}$+ $\lambda$Boo PHL &  & A2IVn+shell (TiII,CaK,HI)    & 175  & 0.155 & 0.64 &  0.1  & --      \\
~~39421 & x &   & A1 Van &   & A2Vp(4481wk)                & 215:          \\
~~47152$^3$ &   &   & & x & A2Vp($\lambda$Boo)          &  25    & 0.212 &  0.77 & 0.1 &  - & \\
~~64491 &   & x & kA3hF0mA3 V $\lambda$Boo (PHL) &  x & A9Vp($\lambda$Boo)          &  15           \\
~~66684A &   &   & B9 Va & x & B9.5Vp($\lambda$Boo)        &  65  & 3.527& 0.73 &  3.5 &  0.9       \\
~~74873 &   & x & kA0.5hA5mA0.5 V $\lambda$Boo NHL &  & A1Vp(4481wk)                &  10           \\
~~75654 & x &   & &  &  ~~~~~~~--                    &             \\
~~79108 &   &   & & x & A0Vp($\lambda$Boo)          &  160          \\
~~81290 & x &   & &  &  ~~~~~~~--                    &             \\
~~83041 & x &   & &  &  ~~~~~~~--                    &             \\
~~84123 & x &   & &  &  ~~~~~~~--                    &             \\
~~84948 & x &   & &  &  ~~~~~~~--                    &             \\
~~87696 &   &   & A7 V & x & A9Vp($\lambda$Boo;met:A5)   & 150           \\
~~89239 &   &   & & x & A2Vp($\lambda$Boo;met:B9.5)  & 135           \\
~~90821 &   &   & kA2hA7mA2 Vn $\lambda$Boo  &  &   &    & &  & & &                    \\
~~91130A &   & x & A0 Va$^{-}$ $\lambda$Boo (PHL)&  x & A0Vp($\lambda$Boo)   & 190    & & & & & M             \\
~~98772 & x &   & & x  & A1IVn                 & 230:               \\
~101108 & x &   & &  & ~~~~~~~--  &                      & 6.7& 3.2          \\
~105058 & x & x & kA1hA7mA1 V $\lambda$Boo (PHL) &  & ~~~~~~~--  & & & & & &  U                                \\
~105759 &   &   & kA2hF0mA2 V ($\lambda$Boo)&                            \\
~106223 & x &   & &  & ~~~~~~~--  &                                \\
~107233 & x & x & kA1hF0mA1 Va $\lambda$Boo PHL &  & ~~~~~~~--  &   &  & & & & U                            \\
~108283 &   &   & A9 IVnp SrII & x & A9Vp($\lambda$Boo)     & 185                \\
~109738  & x &   & &  & ~~~~~~~--              &                   \\
~109980 &   &   & & x & A6Vp($\lambda$Boo)    & 255: & &   & & & M               \\
~110377 &   &   & & x & A6Vp($\lambda$Boo)    & 160                 \\
~110411 & x & x & A0 Va ($\lambda$Boo) NHL &  & A0Vp(4481wk)          & 140   \\
 111604 &   & & & x & A5Vp($\lambda$Boo)          & 180  & & & & & U \\            
 111786 & x & x & A1.5 Va- $\lambda$Boo PHL & x & F0Vp($\lambda$Boo;met:A1)   & 135 \\          
 112097 & &   & & x & F0Vp($\lambda$Boo,met:A7)   &  61 \\            
 118623AB & &   & A7 Vn & x & F0Vp($\lambda$Boo)n         & 190 & & &  1.1 & 1.9 \\            
 120500 &   & x & kA1.5hA5mA1.5 V ($\lambda$Boo) NHL &    &     &  \\                                 
 125162 & x & x & A0 Va $\lambda$Boo NHL & x & A0Vp($\lambda$Boo,met:v.wk) & 110 \\          
 125489 &   &  & & x  & F0 Vp($\lambda$Boo,met:A5)   & 145 \\           
 130158 &   &   & & x & A0IIIp($\lambda$Boo)        &  55  & & & &  & U \\          
 138527 &   &   &  & x & B9.5Vp($\lambda$Boo: Ca,4481 wk)         & ...  & & & & & D\\          
\hline     
                                                                     \end{tabular}
                                                                     \end{table*}
\addtocounter{table}{-1}
\begin{table*}
\caption{The $\lambda$ Boo candidates (continued).}
\label{tab1}
\begin{tabular}{lccclccccccr}
\hline
~~~HD & CC & G & G+GG classif. & AM & AM classif. &AM  &Sep &$\Delta m$ &Sep &$\Delta m$ &Rem.  \\
& &           &      &    &       &   v$ \sin i$ & $''$ & mag & $''$ & mag                  \\
& &           &      &    &       &               & HIP & HIP & WDS & WDS                  \\
\hline
 141851 & x &   & & & A3 Vp(4481wk)n               & 185 &  &  & 0.1&  -\\           
 142703 & x & x & kA1hF0mA1 Va $\lambda$Boo PHL & x & A9Vp($\lambda$Boo)          & 95 \\           
 142994 & x & x & A3 Va $\lambda$Boo PHL & & ~~ ~~~--               & \\                     
 144708$^4$ &   &   & & x & B9Vp($\lambda$Boo)nn & 255: &  & &  3.4 & 3.5 \\                
 149303 & x &   & & & ~~ ~~~--     & \\                               
 153808 &   &   & A0 IV$^{+}$& x & A0IVp($\lambda$Boo)  & 50 &  & &  0.2 & -\\                   
 156954 & x &   & & & ~ ~~~~--          & \\                          
 159082$^5$ &   &   & & x & A0IVp($\lambda$Boo)  & 30  & & & 0.2 & - \\                  
 160928 & x &   & & & ~ ~~~~-- & & & & 0.1 & 0.0     & \\                              
 168740 & x &   & & &  ~ ~~~~--         & &  & & & & M \\                          
 169009 &   &   & & x & A1V p($\lambda$Boo)   &   35 & & & & &  U \\               
 170000$^6$ &   &   & kB9hB9HeA0V(Si)+Note&  x & A0IIIp($\lambda$Boo) & 65 & 0.382 & 1.45 & 0.6 & 1.5 \\                  
 170680 & x & x & A0 Van ($\lambda$Boo) NHL & x & A0Vp($\lambda$Boo)   & 200: \\                
 171948 & x & x &A0 Vb $\lambda$Boo NHL & & ~~ ~~~--                    & \\                
 177120 & x &   & & & ~~ ~~~--              &   & & & 8.0 & 1.5  & U \\                
 177756 &   &   & kB8HeA0IV wk4481+Note & x & B9.5p($\lambda$Boo)n &  ... \\                
 183324 & x & x & A0 Vb $\lambda$Boo NHL & x & A0IVp($\lambda$Boo)  & 105 \\                 
 184190 & x &   & & & ~~~~~--                    & \\                 
 184779 & x & & &   & ~~~~~--                    & \\                 
 192424 & x & & &  & ~~~~~-- & & & & 6.2 & 0.0                   & \\                 
 192640 & x & x & A0.5 Va- $\lambda$Boo PHL & x & A7Vp($\lambda$Boo, met A1,4481 wk)  &  35   & & & & & U  \\           
 193256 & x & x & A2 Va $\lambda$Boo PHL & & ~~ ~~~--     & & &  & & & D  \\                 
 193281$^7$ & x & x & A3mA2 Vb $\lambda$Boo PHL &   & A2.5V & 75 &  & &  4.7 &  3.0 \\                                 
 196821   &   &   & &  x  & A0III($\lambda$Boo)s  &  10 \\            
 198160$^8$ & x & x & A2 Vann $\lambda$Boo PHL:  & & ~~~~~-- & & & &  2.7 & 0.31 \\                                    
 198161 &   & x &  A2 && ~~~~~-- & \\                                    
 204041 & x & x & A1 Vb $\lambda$Boo PHL  & x & A3Vp($ \lambda$Boo)  & 55 \\                  
 210111 & x & x & kA2hA7mA2 Vas $\lambda$Boo PHL &   & ~~~~~--                     & \\             
 212150 &   &   & & x & A0Vp($ \lambda$Boo)   & 180\\                 
 214454 &   &   & A7 IV-V &  x & F0Vp($ \lambda$Boo;met:A6) & 93 \\            
 220061 &   &   & A5 V & x & A5Vp($ \lambda$Boo)  &  135 \\                
 221756 & x & x & A1 Va+ ($ \lambda$Boo) P/NHL &  & A1Vp( 4481wk)         & 75 \\                 
 223352 &   &   & A0Va$^{+}$n & x & A0Vp($ \lambda$Boo)n  & 280: &   & &  3.3 & 7.04 & S \\                
 225218$^9$ &   &   & & x & A3IVp( $\lambda$Boo)s & 20 & & &                     \\ 
\hline     
\end{tabular}

$^1$both A and B are SB,
$^2$different from  Paunzen \& Gray 1997,$^3$occultation and interferometric binary P=22.3yr, 
$^4$A=SB P=4.02d, 
$^5$A=SB P.6.8d,
$^6$A=SB and ${\alpha}$CVn,
$^7$A of a quintuple system C= HD 193256, 
$^8$B=HD 198161, 
$^9$ HIP note:
ambiguous double star solution,
WDS gives Aa sep=$0.1''$ Aa-B sep$=5.2''$ $\Delta m =2.6$; 
B=SB hence the
system is quadruple. 
Remarks from ESA Hipparcos Catalogue (1997)
 D=duplicity-induced variability,
  M=possibly microvariable,
  S=suspected non-single,
  U=unsolved variable.
 \end{table*}

\section{Spectral characteristics}

Abundance analyses have been performed only for a few \lambo stars.
Baschek and Searle (1969) analyzed 5 stars through the curve of growth method and
rejected 2 of them from the \lambo class. Venn and Lambert (1990) made a new
modern analysis of the 3 stars retained as \lambo by the previous authors.
St93 determined the metal abundances for 
15 stars, 2 of which were in common with those considered in the two previous papers.
Heiter et al. (1998) analyzed 3 stars, 2 of which are in common with previous 
authors. Paunzen et al (1999) derived non-LTE  
abundances of C and O for 16 and 22 stars, respectively. 
To these analyses of the visual spectra we can add the semiquantitative study
of the UV spectra of 10 \lambo candidates (3 of which are not included in
our Table 1) made by Baschek and Slettebak (1988). \par
The comparison of the abundances obtained so far shows
the erratic behaviour of the abundance anomalies in different stars:
an emblematic example is the peculiarly slight {\it over-abundance}
of Mg found by St93 in HD 38545 and the almost solar Mg abundance 
found in HD 193281.

Solar abundances of the  elements C,N,O and S is a property that requires
to be further proved; in fact it is essentially based on the Venn and Lambert
(1990) study of 3 stars, but we stress
that the measured lines of these
elements refer to the neutral stage and lie in the red or near infrared
spectral range so that a possible contamination by a cool companion star would
increase the abundances of C, N, O and S.
The other semiquantitative studies by Baschek and Slettebak (1988) and
by Andrillat et al. (1995) are not conclusive, since in these two papers  
opposite
conclusions are drawn in these two papers for what concerns the C,O and S
elements (N does not appear in the Andrillat et al. paper).

The Paunzen et al  (1999) paper strengthens
our remarks on C, N, O and S abundance pattern. In fact,
the mean [C/H] abundance results -0.37 $\pm$ 0.27, 
and the comparison of O
abundances derived previously, show the inconsistency of those derived 
from lines in UV, optical and near-IR regions. 
The extreme cases of HD 141851 and HD 204041
with [C/H]$=-0.81$ pose some doubts on the notion that C is solar
in \lambo stars. 
Moreover,
for HD 204041 even the LTE [C/H]$=-0.75$ 
is not far from the [Fe/H]$=-0.95$ found by St93,
but we recall that a conclusive abundance pattern can be obtained only by 
analyzing all elements with the same model, i.e. with the same parameters 
\teff, log g, microturbulence, vsini and abundances.

The large survey by St93 does not include any abundance determination
of N, O and S; an almost solar abundance is derived for C, but no details
on the measured lines are given; considering the paucity of C lines in the
spectral ranges analyzed by St93, quite possibly most of the results
rely mainly on the CI 4932 \AA~ line.
We note also that a similar, almost solar abundance is derived for the Mg,
which is usually the key element for the \lambo classification, in
one third of his sample stars.
Another peculiar element behaviour is that of Na whose abundance is 
found to be lower than 
the solar one in 5 stars, almost solar in 2 stars and higher than the
solar one in 6 stars.

\begin{figure}
\psfig{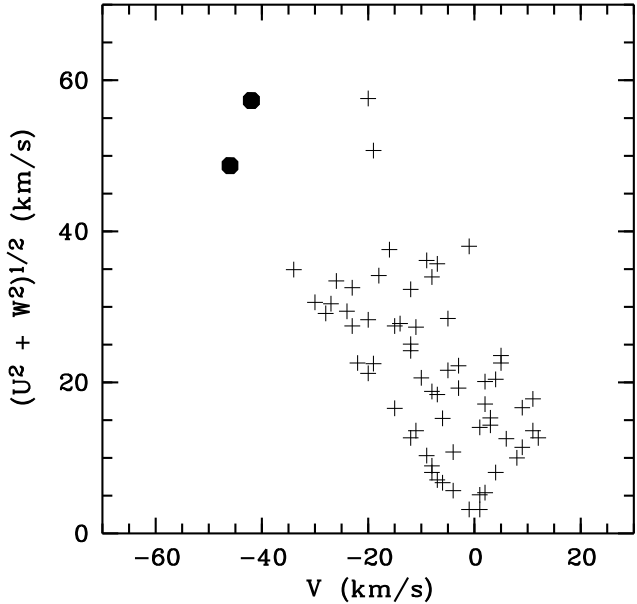}
\caption{ 
Kinematics of \lambo stars. HIP 5321 (HD 6870) and
HIP 47752 (HD 84123) are shown as black dots. 
}
\label{figkin}
\end{figure}

The present knowledge of the Balmer line profiles of the \lambo stars
is based on the studies by Gray (1988)
and  by Iliev and Barzova (1993a, 1993b, 1998).
Gray  pointed out
peculiar Balmer profiles in some $\lambda$ Boo stars
from classification dispersion spectra.
In the spectra of some stars he noted  Balmer lines with
broad wings and weak core, and an inconsistency of the $\beta$ index with
the luminosity class based on the extent of the hydrogen-line wings.
These problems were not present in other $\lambda$ Boo stars .
Thus, according to the appearence of the Balmer lines, he divided the $\lambda$
Boo stars into two distinct classes:
NHL (Normal Hydrogen-Line profile) and PHL (Peculiar Hydrogen-Line profile).
Subsequent
studies by Iliev and Barzova based on high dispersion
photographic spectra confirmed and quantified this peculiarity.
This dicothomy is therefore observationally well established, but
we still lack a theoretical interpretation for it.

\section{Kinematics}

Given the metal-weakness of \lambo stars, it is proper
to wonder whether
some of them are truly metal-poor stars with halo kinematics.
In Table 1 the kinematics of the stars  observed by Hipparcos and with known RV
has been 
computed from  proper motions and radial
velocity as described in Johnson \& Soderblom (1987). 
A left-handed system was 
used in which  $U$ is directed towards the Galactic
anticenter, $V$ in the direction of Galactic rotation and $W$
towards the north Galactic pole.
In Fig. 1 we plot the rotational velocity $V$ versus
$\sqrt{(U^2+W^2)}$, which may be taken as a measure of kinetic energy
not associated to rotation.
All velocities are heliocentric, so that stars with $V$ around
0 $\rm kms^{-1}$ and small $\sqrt{(U^2+W^2)}$ are qualified as disk members.
All the \lambo stars pass this test,
in  agreement with G\'omez et al (1998). This 
outcome was expected
as a consequence of the brightness of the
stars for which velocity data are available; out of
the 14  sample stars  weaker than V=8,
eight were not observed by Hipparcos and 5 lack 
a measured RV. Thus, among the faint stars of our sample,
we could compute the space velocities only for HD 101108.
In Fig.2 we note that 
HIP 5321 (HD 6870) and HIP 47752
(HD 84123), shown as black dots, present a
marginal deviation from disklike kinematics.

The first star has been classified as Population II blue straggler
(Bond \& MacConnell, 1971) and has been identified as a member
of the $\sigma$ Pup group, which is similar to the
globular cluster 47 Tuc 
(Rodgers, 1968, Eggen 1970a, 1970b), which has metallicity
[Fe/H]$=-0.71$ (Da Costa \& Armandroff 1990).
This star also presents a UV excess in the (U-B) colour,
similar to that of Pop II stars, which
is not found in other \lambo candidates.
However, the nature of this star is still matter of
debate.
Paunzen et al (1999) consider it a \lambo star.
Their O abundance of [O/H]$=+0.05$ could be reconciled with
an extreme case of $\alpha$ element enhancement as found
in Pop II stars. However, their C value
[C/H]=+0.14 prompt interpretation as Pop I.
Clearly an elucidation of its nature must rely on 
the determination of the abundaces of C,N,O, $\alpha$ elements
and iron-peak elements using the same stellar atmospheric parameters.
We note here that the abundances of C and O
have been obtained by
Paunzen et al (1999) by assuming  very different values of 
$ v \sin i$ (128 and 
200 km s$^{-1}$, respectively).

HD 84123 
belongs to the cooler (\teff = 6900 K) \lambo candidates selected
only by CC and 
has peculiar
characteristics.
The UBV colours indicate that it does not belong to the MS, as confirmed
by the photometrically derived log g=3.5;
the UV magnitudes 
measured by the TD1 experiment (Thompson et al. 1978)
do not fit those computed with these parameters at 2365 and 2740 \AA.

\section{Atmospheric Parameters}

Photometric indices of the $uvby\beta$ system are examined in order to
determine the atmospheric parameters \teff~ and log\,g from the calibration
of the colour indices by Moon and Dworetsky (1985) (MD).

\subsection{Colour excess}

\begin{figure}
\psfig{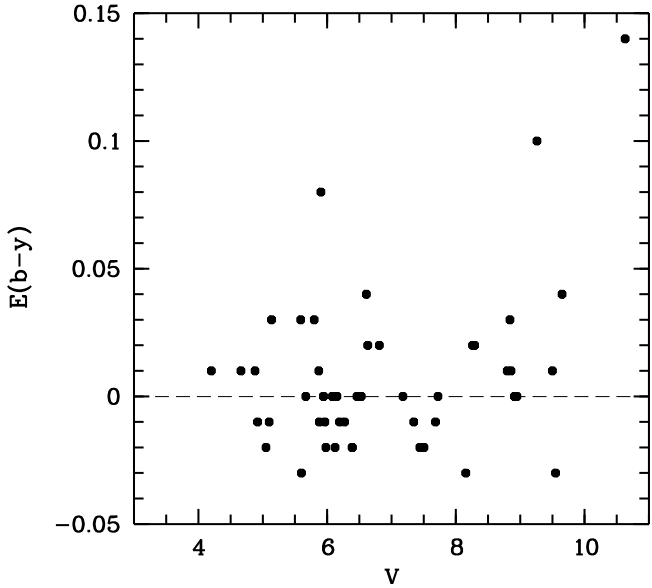}
\caption{The colour excess E(b-y) of \lambo stars as a function of 
their apparent magnitude.}
\label{fig1}
\end{figure}

The observed indices are taken from the catalogue of Mermilliod et al.
(1997).
For 5 stars ( HD 5789, HD 171948, HD 177120, HD 192424 and HD 198161)
they are 
not available and for HD 184190 only the $\beta$ index was 
measured; 
for the remaining 82 stars the colour were dereddened using the
UVBYLIST
code of Moon (1985).
In fact, 30 stars in Table 1 have a visual magnitude
weaker than 6.5 so that we cannot neglect a possible
influence of reddenning.
In 36 \% of the stars the colour excess has a negative value, which is
twice the occurrence found in the sample of 71 bright dwarf A0 stars
recently analyzed by Gerbaldi et al (1999) with the same procedure.
The most negative
colour excess among this latter sample was of -0.02, found for one star
only, while for the other stars with negative colour eexcess 
it was equal to -0.01.
Among the objects of the present sample the stars with
negative colour excess equal to or less than -0.02 are at least 10.
The UVBYLIST code requires that the stars are assigned to one of six
possible groups, depending on spectral type and colours; for several stars
the choice of the  appropriate group
is ambiguous owing to
inconsistencies of the colour indices and spectral type.
We decided to give
less importance to the spectral type and
rely more on the unreddened
$\beta$ index. For the still remaining doubtful
cases an 
unambiguous choice between the parameters derived with different
choices 
cannot be performed without further information; for the most doubtful
cases the various possible choices have been retained.

Since the MD 
procedure has been established for solar abundance stars, we
investigated 
if the derived colour excess is related to the lower than solar
abundances, attributed to these stars by analyzing, with the same procedure,
the theoretical colour indices computed for [M/H] = -~0.5 and
-~1.0 (Castelli 1998).
The colour indices computed for abundances 10 times lower than solar
produce a spurious  colour excess up to 0.03 for \teff=8000 K,
but never a negative value.
We thus conclude that a negative value of E(b-y)
cannot be  directly related to metal underabundances.

Strange enough
the value of E(b-y) is not correlated to the
V magnitude which, for these non evolved stars, is expected to be roughly
related to their distance (Fig. 2).

A high value of v\,sin\,i is expected to affect the photometric (Collins \& 
Smith 1985) and spectroscopic (Cranmer \& Collins 1993; Collins \& Truax 1995)
characteristics of A-type stars. The inspection of the stars for which AM measured
the v\,sin\,i has not revealed any correlation between the anomalous negative
colour excess and the v\,sin\,i value.

We observe also that for 12 stars the photometric measures refer to two
components of a binary system and thus require some sort of disentangling 
from the influence of the companion.

\subsection{\teff~and log\,g}

The parameters \teff~ and log\,g have been derived from the MD TEFFLOGG code
according to the above described choice of colour excess. 
The table of these parameters is available from the authors.

The stars cover a broad domain of the HR diagram around the prototype
$\lambda$ Boo; in fact \teff~ spans from 6500 (HD 4158 and HD 106223) to 14500 K
(HD 22470) and log\,g from 2.87 (HD 108283) to 4.50 (HD 294253). We note that
the 8 stars with \teff~ lower than 7100 K belong to the CC list and the 5
stars with \teff~ higher than 10500 K are from AM; a systematic shift in 
\teff~ probably exists between these two sources. However, 
the bulk of the stars
in the two papers lies in the temperature range they share.

We conclude that before  proceeding further with the discussion of
the region of the HR diagram covered by these stars it is necessary to
examine their photometric properties and their peculiarities.

\subsection{Colour-magnitude diagram}
\begin{figure}
\psfig{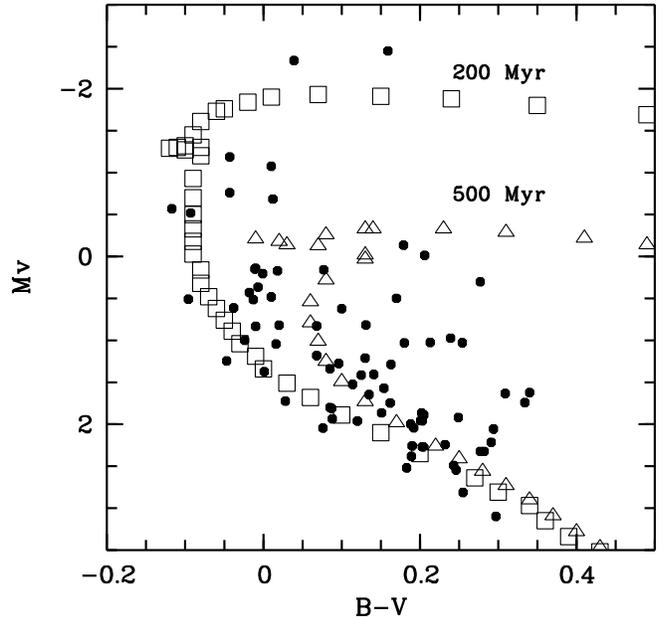}
\caption{The colour magnitude diagram: dots indicate the \lambo stars; squares 
and triangles the isochrones of Bertelli et al (1994) for 200 and 500 Myr 
respectively.}
\label{fig2}
\end{figure}

For the stars with a Hipparcos measured parallax we may readily
build a colour-magnitude diagram. 
Undereddened colours are used, the only ``faint'' star being HD 105058, which is
expected to be unreddened (Faraggiana et al 1990).
This is displayed in
Fig. 3, where the isochrones of solar composition and
ages of 200 Myr and 500 Myr of the Padova group
(Bertelli et al. 1994) are also shown.
We note the inconsistency between the position in the HR diagram
and the log g values derived from $uvby\beta$ photometry and
by using the MD procedure; this point deserves further analysis.
Two facts are clear from this comparison:
1) most of the \lambo stars lie outside the main sequence;
2)  a range in ages between 0.2 and 1 Gyr is necessary
to explain the observed dispersion in the colour
magnitude diagram, if we consider these stars in  post-main sequence phase.
The alternative hypothesis that {\it all} these stars are very young 
objects in the
late phase of their pre-main sequence evolution is highly improbable;
in fact, according to the scenario proposed by Turcotte and Charbonneau
(1993) the accretion of gas, but not of dust, should occur in
less than 10$^{6}$ years and the number of bright \lambo candidates
would imply that the star formation process is still very active in the
solar neighbourhood,
which would imply a large
number of MS B stars in a similar volume.
A fraction of pre-main sequence stars cannot either be excluded or demonstrated
on the basis of only these data.

\section{Peculiarities of \lambo candidates}
We performed a systematic search of known binaries among the \lambo 
candidates of Table 1. \par
HD 38545 and HD 141851 are visual binaries whose speckle
interferometric measurements are given by McAlister et al (1993). 
Given the small separation of the pairs, ground-based spectra
will always be the combined spectrum of these binaries; for the first star,
Hipparcos data demonstrate that the effect 
of the companion is surely not negligible.
Starting from the analysis of the H$_{\gamma}$
profile, Faraggiana et al (1997) were able to show that a
third star, HD 111786, is indeed a binary system and
this finding was supported by the identification of the lines
(in the visual, but not in the UV range below 2000 \AA) of the cooler 
companion, which
are narrow if compared to those of the primary star, this happens
quite likely because the companion is seen pole on.
In order to determine abundances,
the spectra of the two components must be disentangled.

Twelve out of 89 stars of our sample are known to be visual
binaries according to the Mermilliod et al. (1997) catalogue.
For these binaries only the combined colour indices have been measured 
and therefore they cannot be safely
used to derive the atmospheric parameters of the
primary component. In order to evaluate the influence of the companion
star, we extracted the angular separation and the magnitudes for the
components A and B from the Washington Visual Double Star Catalog (Worley
\& Douglass 1997; hereafter WDS); these data are given 
in columns 10 and 11 of Table 1 for the
objects with an angular separation of less than 10 arcsec.
The contamination on the observed spectra
depends on the luminosity difference of the components and on the slit
width of the spectrograph on the sky; for several stars the observed spectra
are expected to be affected by the companion star and, in some cases
(HD 290492, HD 38545, HD 47152, HD 141851, HD 153808, HD 159082, HD 160928,
HD 170000 and HD 225218), a composite spectrum
cannot be avoided with observations from ground instruments unless it can be
demonstrated that the companion
luminosity is much weaker. 

The effect of the secondary star should underlie the discordant
classifications  proposed for HD 225218, either B9 III or A3 Vs;
moreover its UV magnitudes, as determined by the TD1 experiment
(Thompson et al. 1978), would suggest a much lower reddening than that
derived from the Str\" omgren photometry, i.e.
E(b-y)=0.03, a value more coherent with
the stellar visual magnitude.

The two stars HD 141851 and HD 149303, being X-ray sources (H\"unsch et al. 
1998), are expected to have cool companions.

Significant discrepancies in the magnitude and
photometric colours
of the AB system HD 193281 are reported in the literature (BSC).

The ESA Hipparcos Catalogue (1997) allowed to complete information on
some stars and to add new binaries; these data are collected in 
columns 8 and 9 of Table 1.

Further known binaries are:

$\bullet$ The already quoted spectroscopic binary HD 111786 and 
HD 84948 and HD 171948. The 4 components of the last two binaries are 
all \lambo stars, according to Paunzen et al (1998); however, we note that the
Mg does not show underabundance  higher than the other metals and that the 
abundances of the key elements C,N,O and S are not given.

$\bullet$ HD 142703 is a suspected occultation double (BSC).

$\bullet$ HD 79108 is a suspected SB (BSC), but we could
not retrive any further information on the possible influence of the companion
on photometric and spectroscopic data.

$\bullet$ The oxygen spectrum indicates that HD 149303 is an SB system
(Paunzen et al 1999).

We also note the inconsistent classifications assigned to HD 22470 i.e. a \lambo
star (AM) or He-weak with variable intensity of the SiII 4128-30 doublet
(Gray 1988). The explanation of the peculiarities has been given by
the Hipparcos detection of its duplicity (see columns 8 and 9 of Table 1). 
Similar remarks apply 
to two other binaries, HD 47152 and HD 170000, which have been both classified,
as single objects, either Ap or \lambo. \par
Two more spurious \lambo candidates are HD 130158 which is in reality
an Ap Si-$\lambda$4200 star (Gray 1988) and HD 159082 which is a
B9 Hg-Mn (see, for example, Hubrig \& Mathys 1996).

The peculiarity of HD 108283, which has a very high c$_1$ value and
therefore a
derived log\,g which is the lowest of our sample stars, but also a very high
v sini value,  deserves further
analysis before being assigned to the \lambo class; Hauck et al. (1998)
rejected it from the \lambo class.

We conclude that from ground and space observations close duplicity which is
able to affect
the observed spectrum,  has been already
observed or suspected for 24 $\%$ or 33 $\%$ (if the stars
classified U by Hipparcos are included) of the stars of Table 1 and that some spurious 
\lambo candidates are present.

\section{Predicted composite spectra}

In the abundance analyses $\lambda$ Boo stars have been
generally considered as single stars.
The exception is the analysis of the two SB2 stars HD 84948 and 
HD 171948 by Paunzen et al (1998).
In the present section we consider the influence of a possible companion
on model parameters and on derived abundances and discuss the ability
to pick out the already known binaries on the basis of spectra.
This has a direct bearing on the issue whether binarity may lead
to significant systematic errors in the abundance analysis and
classification of $\lambda$ Boo stars.

In a binary system, if the angular separation of the two components is
too small to be 
detected, and the two stars have a 
similar mean to high projected
rotational velocity,  the spectral lines of the components are
not resolved in the observed spectrum and duplicity is not easily
detected spectroscopically. If the two components have significantly
different parameters, 
the duplicity should appear when a large spectral range
is covered by observations, 
the contribution of each star being different
at different wavelengths.
This implies that there are doubts on binarity, 
the analysis of spectral ranges of only a few hundred
\AA~ may not be sufficient to establish or reject
the binarity.

If a binary star is analyzed 
in the classical  3500-5500 \AA~ range,
the average 
values of \teff~ and log\,g are derived from the combined light;
synthetic spectra are computed from the adopted model atmosphere and the
best fit with observations is obtained by adjusting the microturbulence
value and the chemical abundances.

\begin{figure}
\psfig{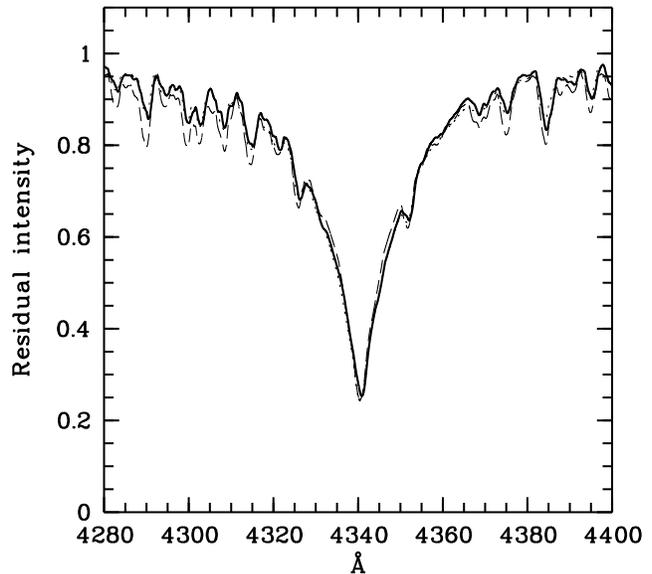}
\caption{The comparison between a computed composite spectrum and a
single one in the region of H$_{\gamma}$. 
The composite (thick line) consists of two spectra
corresponding to \teff = 7200 and 9000 K,
log\,g =4.0, vsini=50 km s$^{-1}$, luminosities L/L$_{tot}$ =0.2 and 0.8 
respectively, and solar abundances; 
the two spectra have been slightly shifted
with respect to each other.
The two single spectra with which the composite is compared
correspond to
\teff = 8000 K, log\,g =4.0, solar abundances 
and  \teff = 8250 K log\,g =4.0 and [M/H]=-0.5, both
with vsini=100 km s$^{-1}$.
}
\label{fig5}
\end{figure}

\begin{figure}
\psfig{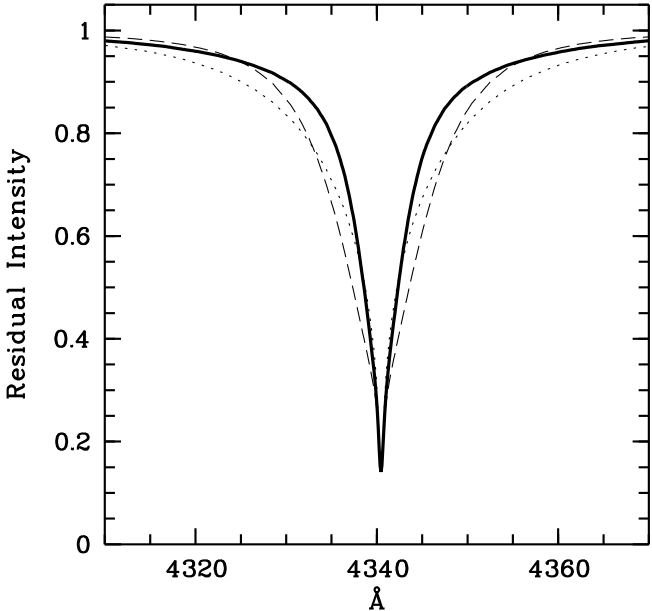}
\caption{Example of how a PHL profile (thick line)
may be created
by combining two single spectra of \teff = 9500 K, log g = 2.0
and \teff = 8500 K, log g =4.0, and luminosity ratios
$L/L_{tot}$ of 0.8 and 0.2 respectively with a
single one in the region of H$_{\gamma}$. 
The dashed profile corresponds to \teff = 7000 K, log g =4.0 and fits the
wings. The dotted profile corresponds to \teff = 9750 K, log g = 3.0 and
fits the core. 
This is  an example  to illustrate  the effect  which 
is much larger than what is
actually observed in \lambo PHL stars.
}
\label{fig5}
\end{figure}

In Fig. 4 we illustrate the result of the comparison of a composite spectrum
with two single-star spectra.
We note that in these spectra the H$_{\gamma}$ profile is practically 
the same, but most of the metal lines 
simulate a metal underabundance
in the composite spectrum.
The hydrogen profile, in this case, would
be considered as NHL. 
Metal underabundances would be obtained from the combined spectrum if
analyzed by disregarding its binary nature.
A similar result on the apparent metal underabundances of the 
Am triple system,
$\pi$ Sgr, if analyzed as a single object, has been obtained  by Lyubimkov \&
Samedov (1987). 

To illustrate how a PHL profile can be produced by the combination 
of 2 single profiles, we constructed the composite H$_{\gamma}$ line given
in Fig. 5 
and compared it with
the single spectra computed with \teff=7000 K, log g=4.0, which fits the wings, 
and with \teff=9750 K, log g=3.0, which fits the core of the composite spectrum.
All these profiles refer to solar abundance models.
The example has been constructed in such a way as to
enhance at most the effect which is observed in PHL stars
and is not intended to simulate any really observed object. 

\section{Discussion}

In the previous sections we stressed the fact that the $\lambda$ Boo
candidates of Table 1 constitute a non homogeneous group.
By adding the informations spread in the literature, we summarize that
among \lambo stars: \par
- some but not all show PHL profiles (Gray 1988, 1998); \par
- some but not all have an IR excess (2 stars in Sadakane \& Nishida 1986;
2 stars in  Cheng et al 1992; 1 star in Grady et al 1996; 2 more in 
King, 1994); \par
- some but not all have a shell surrounding the star which is detected by the
presence of narrow circumstellar components of Ca II K line (Holweger and
Rentzsch-Holm 1995) and of metal lines (Hauck et al 1995 and
1998; Andrillat et al 1995); \par
- some but not all show the UV broad absorption feature centered
on $\lambda$1600 \AA~ (Baschek et al 1984; Faraggiana et al 1990; 
Holweger et al 1994). We recall that the combined effect of the stellar
flux drop and the lowering IUE sensitivity is responsible for the mostly
underexposed IUE spectra of the middle and late A-type stars, so that the 
$\lambda$1600 \AA~ could be searched only among the hottest \lambo
candidates; \par
- for three of them an abundance pattern similar to that of the ISM has
been derived (Venn \& Lambert 1990).\\

On the basis of the few abundance coherent analyses available, 
several hypotheses
have been made on the age of these  stars:

i) very young stars which have not reached the main sequence (Waters et al
1992; Gerbaldi et al 1993;  Holweger \& Rentzsch-Holm 1995).

ii) dwarfs in the middle of their life on the
main sequence, with an age of  10$^7$-10$^9$ years (Iliev \&
Barzova, 1995);

iii) quite old objects representing a merger of binaries of W UMa type
(Andrievsky 1997).\\

The only property common to all $\lambda$ Boo stars of Table 1
is the weakness of most metal lines, which is also confirmed by the negative
$\Delta$a values (a photometric index measuring the blanketing in the region
$\lambda$5200 \AA) measured by Maitzen \& Pavlovski (1989a and 1989b)
for the stars they observed.

Taking this common characteristic of the group as a starting point,
we inspect the possible causes that may produce a weak-lined spectrum. \\

The classical explanation of metal underabundances, for stars belonging to
Population I, is related to the existence of single
stars with peculiar atmospheres in which some elements are depleted by
different amounts. 

Disturbing is that the abundance pattern is not the same in the
$\lambda$ Boo candidates analyzed up to now; one particular element shows
different abundance peculiarities in different stars, so that
it is difficult, at present, to establish the average chemical composition of
\lambo stars and thus to elaborate a theory explaining 
the phenomenon.

Venn \& Lambert (1990) formulated the hypothesis that the \lambo
phenomenon could be the result of accretion of gas but not of dust from
circumstellar or interstellar material.
The only modern and detailed analyses available at present
are the two papers by Venn \& Lambert (1990) and by St93.
We compared the metal abundances derived for the $\lambda$ Boo stars
by these authors with those of the ISM as given by Savage \& Sembach (1996).
This comparison shows that the similarity is only marginal;
in the first place there is a large difference from star to star  in the
abundance of any given element, unlike what happens in the ISM,
in the second place
the highest underabundances in the ISM are those of Ca and Ti while in most
$\lambda$  Boo candidates it is that of Mg.\\

Among possible explanations of the \lambo phenomenon
one should also take into account the possibility that these stars
indeed belong to a metal-weak population.
The kinematic data imply that \lambo stars belong
to the disk population, which
has indeed a metal-poor population
in the range -0.5$<$[Fe/H]$<$-1.0 and possibly the
thick disk has a metal-weak tail at much lower metallicities
(Beers \& Sommer-Larsen 1995).
We must examine the two possible cases that
\lambo stars are either Main Sequence (MS) and therefore relatively young,
or on the Horizontal Branch and therefore old.

The main argument to reject the hypothesis that \lambo stars
are metal-poor MS stars   is that
while in \lambo stars Mg, Si,  Ca and Ti   are among the most
underabundant elements,
in metal-poor stars  the even-Z light elements,
synthesized by
$\alpha$ capture processes,  show an
increasing  enhancement over iron with decreasing metallicity,
reaching a 0.4 - 0.5 dex enhancement at [Fe/H]$=-1.0$.

To distinguish  Blue Horizontal Branch  (BHB) stars
from
$\lambda$ Boo stars on the basis of spectroscopic properties alone,
is not trivial.
However, the BHB population in the solar neighbourhood
would be uncomfortably large if most \lambo belonged to this
class.
A further argument against the 
BHB hypothesis is that most \lambo stars
are characterized by mean to high projected rotational velocities, while
BHB stars are all slow rotators. 
All fast rotators may be thus rejected as BHB stars.
Although the possibility could be still considered open
for slowly rotating \lambo stars
(e.g. HD 64491 and HD 74873
(Paunzen \& Gray (1997) 
and a few others proposed  by AM), their number is very
small.
We recall that low v$\sin i$ stars
may be either intrinsically slow rotators
or fast rotators seen at high inclination.

From the foregoing discussion
we reject the hypothesis of membership of the class to a metal-poor 
population and do not discuss it any further.

\bigskip

A completely different origin of weak metal lines is that produced
by stellar duplicity.
Examples of how a composite spectrum, which is the average of two actual components of
not very dissimilar spectral type, can be classified as Mg-weak is given by
Corbally (1987) for HD 27657 and HD 53921.
Corbally remarks also that the AB spectrum of HD 41628 "is close to imitating
a $\lambda$ Boo star, but the A5 Balmer line class is a compromise between
A7 V strength and A3 V wings... example of two normal parent spectra producing
a peculiar composite".

The effect of veiling in the spectrum of a binary with 
components not very dissimilar from one another (M=2 and 1.4 solar masses) has been investigated in detail by
Lyubimkov (1992). Most of his analysis, devoted to Am stars, refers
to composite spectra (computed for 4 selected evolutionary phases) obtained by
combining two spectra for which solar abundances are adopted only for
elements lighter than Ti. A general apparent underabundance of these elements
is derived by his computations when the original duplicity is neglected,
in agreement with the weak metal lines obtained by our example plotted in 
Fig. 4.

According to the data collected in the previous sections, 11 stars of our
original sample are doubles with an
angular separation smaller than 1.2 arcsec, 3 stars are SB2 and 4 are 
probably non-single, according to the Hipparcos data.
In conclusion,  for 18/89= 20 $ \%$ of our sample stars duplicity 
must be examined in further detail before determining
atmospheric abundances.

Grenier et al. (1999)
in their radial velocity study of a sample of B to F stars, 
included in the
Hipparcos catalogue, 
obtained spectra for 16 stars of our sample.
Of these 12
are suspected, probable or established binaries,
only 4 of these are among the 18 known binaries previously mentioned.
If all of them will be confirmed to be binaries the percentage
will raise to 29\% .
We note also that
11 of these stars are in 
common with the G list and 8 are classified as PHL by him.

If we apply the present knowledge to the 15 stars analyzed by St93,
we see that 2 of them are SB2 (HD 38545 and HD 111786), for
HD 198160 and HD 198161 the atmospheric parameters, derived from the
combined photometric indices, require the hypothesis that the two stars are
strictly the same so that the same \teff~ and log\,g can be adopted.
The duplicity of these stars
requires to be further examined in order to determine accurate single
elements abundances. Furthermore, the variability of the 5 variable stars 
must be examined to assess that its amplitude does not affect the 
photometrically derived atmospheric parameters.

High S/N spectroscopic data of spectral regions in which not severely blended
features are present are necessary to discriminate between "veiling"
(spectral lines when they retain the breadth of their temperature type,
but are shallower than normal (Corbally 1987)) which indicates a composite 
spectrum and normal
profiles with weak intensities, which are sign of real metal underabundances.
Such discrimination,  however, 
becomes extremely difficult when the observed spectrum
is characterized by broad and weak metal lines as in most \lambo candidates.

What we can expect in a composite spectrum of two similar A-type stars are
Balmer lines broader than those of the single components by an amount which
depends on the relative RV of the components, so simulating a star with a
higher log\,g value when compared to computed spectra or intrinsically very
high for an early A-type star as it may be the case of HD 294253 (the
parameters derived by MD programs are \teff=10370 K log\,g=4.50).
Moreover, the composite Balmer line
profile will present a flat inner core which depends on the difference of the
two radial velocities as well as a global profile which may be different from what
is expected from the dominating broadenings: Doppler
core and linear Stark wings.

For the stars recognized to be double by speckle observations, and not
observed by the Hipparcos satellite, the extraction of luminosity ratios
from speckle data will be fundamental to better define the character of the
two components.
Algorithms to extract luminosity ratios from speckle data have been
developed, but these techniques are still limited (Sowell \& Wilson
1993).
If the luminosity of the companion is large enough (of the order
of 25\% of the total luminosity), the veiling may not be neglected;
in fact the metallic lines will appear weaker, thus
leading to an underestimate of the metallicity.

The  IR colours could also prove to be powerful diagnostic tool for
the presence of cooler companions.
A cool companion of HD 111786 was predicted by its photometry in the
J,H and K bands by Gerbaldi (1990) on the basis of the discrepancy with the
(B-V) value
and was ascribed to a probable  cool companion.

The foregoing discussion leads us to formulate the hypothesis
that a considerable fraction of \lambo candidates
are in fact
binaries. This is supported by the large fraction of
binaries recently discovered among \lambo stars either
through the speckle technique or by the Hipparcos experiment.
Also the high number of stars with a "blue" colour excess supports that our
binarity hypothesis  is at the origin of distorted energy distributions 
and of not coherent uvby$\beta$ indices of several
\lambo candidates.
The PHL phenomenon cannot be easily explained if the stars
are single, however its explanation becomes trivial
if the stars are binary,  as has been
demonstrated in the case of 
the stars HD 38545 and HD 111786, classified PHL by Gray (1988, 1998),
which indeed turned out to be binaries.
The apparently erratic abundance patterns pose serious problems
to the accretion hypothesis, but again it may be easily reconciled
in the case of binary stars.

The fact that some of the stars are binaries does not exclude
the possibility that chemical peculiarities are actually present
in their atmospheres. However their quantification requires
that the binarity is properly accounted for.

\bigskip
\section{Conclusion}
We have shown that
each author has his own definition  and list of $\lambda$ Boo candidates
and these lists only partly overlap. Until all the classification
schemes converge into a single  with a physical basis there is little
hope of understanding the \lambo phenomenon. 

Recent observations, mainly by speckle interferometry and by the Hipparcos
satellite, have detected the binary nature of several \lambo candidates
and other candidates  have been recognized on the basis of high resolution spectra.

We make the hypothesis that abundance anomalies are, at least partly, due
to the effect of veiling in a composite spectrum and that other still
undetected binaries are likely to be present among the objects collected in 
Table 1.
Distorted and uncertain colours (e.g. stars with negative 
colour excess and reddened bright stars) and peculiar Balmer line profiles 
are reasons for suspecting duplicity.

The photometrically derived atmospheric parameters of close visual binaries
refer to the average photometric indices
of the components and an abundance analysis based on them requires that
the two components form a twin pair or have very different luminosities.
Moreover, for some of these binaries the angular separation is
such that a composite spectrum cannot be avoided.

\begin{acknowledgements}
We thank Fiorella Castelli for helpful discussions and suggestions and
for having put at our
disposal the update version of the Kurucz BINARY program.
Use was made of the SIMBAD data base, operated at the CDS,
Strasbourg, France.
Grants from MURST 40$\%$ and 60$\%$ are acknowledged.

\end{acknowledgements}

\end{document}